\documentclass{ws-procs975x65}

\begin{document}

\title{
Measuring neutrino masses with supernova neutrinos
\footnote{\uppercase{W}ork done in collaboration with \uppercase{J.I. Z}uluaga, 
and partially supported by \uppercase{C}olciencias under contract 1115-05-13809.}}

\author{E.~Nardi}

\address{
 Laboratori Nazionali di Frascati, C.P. 13, I00044 Frascati,  Italy \\
Instituto de F\'\i sica,  Universidad de Antioquia, A.A.1226, Medell\'\i n, Colombia\\
E-mail: enrico.nardi@lnf.infn.it}

\maketitle

\abstracts{A new method to study the effects of neutrino masses on a
  supernova neutrino signal is proposed.  The method relies exclusively on the
  analysis of the full statistics of neutrino events, it is independent of
  astrophysical assumptions, and does not require the observation of any
  additional phenomenon to trace possible delays in the neutrino arrival
  times.  A statistics of several thousands of events as could be collected by 
  SuperKamiokande, would allow to explore a neutrino mass range somewhat below $1\,$eV.}

\section{Introduction}

Already long time ago it was realized that Supernova (SN) neutrinos can
provide valuable informations on the neutrino masses.\cite{Zatsepin:1968}
The basic idea relies on the time-of-flight delay $\Delta t$ that a neutrino of
mass $m_\nu$ and energy $E_\nu$ traveling a distance $L$ would suffer with respect
to a massless particle:
\begin{equation} \label{delay}
\frac{\Delta t}{L} = \frac{1}{v} - 1 
\approx 
\left(
\frac{5.1\,\rm ms}{10\,\rm kpc}\right)  
\! \left(\frac{10\, \rm
    MeV}{E_\nu}\right)^2  \!\! 
  \! \left(\frac{m_\nu}{1\,\rm eV}\right)^2\,,
\end{equation}
where for ultra-relativistic neutrinos $1/v=E_\nu/p_\nu \simeq 1+ m^2_\nu/2 E_\nu^2$
was used.  The dispersion in the arrival time of about twenty $\bar \nu_e$ from
supernova SN1987A was used in the past to set the model independent limit
$m_{\bar \nu_e}<30\,$eV,\cite{Schramm:1987ra} while a recent detailed reanalysis
obtained, within the SN delayed explosion scenario, $m_{\bar
  \nu_e}<5.7\,$eV.\cite{Loredo:2001rx} Since SN1987A, several efforts have been
carried out to improve the sensitivity of the method, while awaiting for the
next Galactic SN explosion.  Often, these approaches rely on ``timing'' events
related to the collapse of the star core, that are used as benchmarks for
measuring the neutrino delays.  The emission of gravitational
waves,\cite{gravit1,gravit2} the $\nu_e$ neutronization burst,\cite{gravit2} the
initial steep raise of the neutrino luminosity,\cite{steep} and the abrupt
interruption of the neutrino signal due to a further collapse into a black
hole\cite{blackhole} have been used to this aim.  However, there are some
drawbacks to these methods: firstly only neutrinos with arrival time close to
the benchmarks are used; secondly the observation of the benchmark events is
not always certain, and in any case some model dependence on the details of
the SN explosion is generally introduced.  A new method that is free from
these drawbacks was proposed in Ref.~\refcite{Nardi:2003pr}. The method relies
only on the measurement of the neutrinos energies and arrival times and uses
the full statistics of the detected signal. It is also remarkably independent of particular
astrophysical assumptions since no use is made of benchmarks events.  The
basic idea is the following: in the idealized case of vanishing experimental
errors in the determination of the neutrino energies and arrival times, and
assuming an arbitrarily large statistics and a perfectly black-body neutrino
spectrum, one could use the events with energy above some suitable value $E^*$
(to suppress the mass effects) to reconstruct very precisely the evolution in
time of the neutrino flux and spectrum.  Once the time dependence of the full
signal is pinned down, the only parameter left to reconcile the time
distribution of the low energy neutrinos with the high energy part of the
signal would be the neutrino mass, that could then be nailed to its true
value.  Of course, none of the previous conditions is actually fulfilled.  In
 \v{C}erenkov detectors as SuperKamiokande (SK) the uncertainty in the
energy measurement is important and must be properly taken in to account. The
statistics is large but finite, and finally, the SN neutrino spectrum is not
perfectly thermal.\cite{neutrino-spectrum} Nevertheless, a good sensitivity to
the mass survives, allowing to disentangle
with a good confidence the  case $m_\nu =0$ from  $m_\nu=1\,$eV.

The idea outlined above was tested in  Ref.~\refcite{Nardi:2003pr}
by proceeding in two steps:
{\it i)}~First a set of synthetic neutrino signals is generated, according to
a suitable SN model.\cite{woosley} Neutrinos are then propagated from the SN to the
detector assuming two different mass values: $m_\nu = 0\,$ and $ 1\,$eV.
{\it ii)}~The detected signals are then analyzed with the aim of disentangling
the two cases $m_\nu = 0\,$ and $1\,$eV.  Only the
SN-Earth distance is assumed to be known, while all other
quantities, like the spectral functions and the detailed time evolution
of the neutrino flux are inferred directly from the data.

For the time evolution of the neutrino flux and average energy the results of
the SN explosion simulations given in Ref.~\refcite{woosley} were used. The
neutrino energy distribution was modeled by a Fermi-Dirac (FD) spectrum with
time dependent spectral temperature $\hat T(t)$ and a `pinching' factor
$\hat \eta(t)$ introduced to account for spectral
distortions\cite{neutrino-spectrum}.  Each neutrino is labeled by its
emission time $t_\nu$ and by its energy $E_{\nu}$, and the corresponding positron
produced through the reaction $\bar \nu_e\, p\to e^+\, n $ is also identified by a
pair of values $(E^e, t^e)$ generated by taking into account the detection
cross section $\sigma(E_\nu)$ and the SK energy threshold and resolution.

The synthetic signals are then analyzed by means of the likelihood function
\begin{equation}
{\mathcal L} =  S\big(\epsilon;T(t),\eta(t)\big) \times \Phi(t+\delta t;b,d,f)\times
\sigma(\epsilon) \,,
\end{equation}
where $\epsilon$ is the neutrino energy inferred from the positron energy.  The
effective temperature $T(t)$ and pinching $\eta(t)$ of the FD spectral
function $S$, and the detailed shape of the parametric function $\Phi(t+\delta t;b,d,f)$
that describes the neutrino flux are directly derived from
the data.\cite{Nardi:2003pr} Given a value of $m_\nu$ the time delay of each
neutrino is computed according to its energy $\epsilon_i$, and subtracted from its
arrival time $t_i$. For the new array of times $-\log{\mathcal L}$ is then
evaluated, and minimized with respect to the flux shape parameters, until the
value of the mass corresponding  to the absolute minimum is found.  The power 
of the method for disentangling $\hat m_\nu=1\,$eV from $\hat m_\nu=0$ was studied
by assuming in turn the two energy thresholds $E_{\rm tr}=5\,$ and
10$\,$MeV, and two SN-Earth distances $L=10\,$ and $20\,$kpc. For each case
40 samples were analyzed.  Table~1 summarizes some of the results. The first
raw gives the percentage of times in which the 95\% c.l. lower limit $m^l_\nu$
is {\it larger} than the input mass $\hat m_\nu$.  The second raw refers to the
cases when the upper limit $m^u_\nu$ is {\it smaller} than $\hat m_\nu$.  Number
in parentheses correspond to $L=20\,$kpc.  These figures characterize the
percentage of `failures' of the method, that therefore appears to be reliable
in about 90\%-95\% of the cases.  The third row gives the percentage of times
when $m_\nu=0\,$ is excluded at 95\% c.l. when the signal is generated with
$\hat m_\nu=1\,$eV. We see that for  $E_{\rm tr}=5\,$MeV and $L=10\,$kpc the method is
successful in more than 50\% of the cases.
%
\begin{table}[bt]
\tbl{Results of the fits to the neutrino mass.}
{
\begin{tabular}{@{}ccccc@{}}
\hline
{} &{} &{} &{} &{}\\[-1.5ex]
$E_{{\rm tr}}$:
&\multicolumn{2}{c}{5\,MeV}  
  &\multicolumn{2}{c}{10\,MeV}  \\[1ex]
$\hat{m}_\nu$:     & 0 eV    & 1 eV     & 0 eV      & 1 eV  \\[1ex]
\hline
{} &{} &{} &{} &{}\\[-1.5ex]
$m_\nu^l > \hat{m}_\nu\,$(\%)& 5 (11)& 4 \ (5) & 5 (10)& \ 11 (5)\\[1ex]
$m_\nu^u < \hat{m}_\nu\,$(\%)& 9 (6) & 2 \ (5) & 12 (6)& \ 10 (7) \\ [1ex]
$ m_\nu^l > 0\>$ \  (\%)   & --    & 55 (40) & --    & 28 (23)\\[1ex]
\hline
\end{tabular}\label{table} }
\end{table}
%
The results in table \ref{table} were derived relying on two main simplifying
approximations: {\it 1)}~the neutrino energies were generated assuming a
`pinched' FD spectrum and fitted with a similar two-parameters energy
distribution; {\it 2)}~no effects of the neutrino oscillations were taken into 
account.  The effects of these approximations have been analyzed in
Ref.~\refcite{Nardi:2003pr} by carrying out 40 simulations using the
{\it numerical} spectra given in Ref.~\refcite{neutrino-spectrum} and assuming
a mixed $\bar \nu_e$--$\bar \nu_\mu$ composite spectrum as would result from
neutrino oscillations.  It was found that even when the $\bar \nu_e$--$\bar
\nu_\mu$ spectral differences are assumed to be unrealistically large, the
sensitivity of the method to the neutrino mass is only mildly affected.
%


\begin{thebibliography}{99}

\bibitem{Zatsepin:1968}
G.T. Zatsepin, {\it JETP Lett.\ } {\bf 8}, 205 (1968).  

\bibitem{Schramm:1987ra}
D.~N.~Schramm,
 {\it  Comments Nucl.\ Part.\ Phys.\ }  {\bf 17}, 239 (1987).

\bibitem{Loredo:2001rx}
T.~J.~Loredo and D.~Q.~Lamb,
 {\it  Phys.\ Rev.\ } {\bf D65}, 063002 (2002). 

\bibitem{gravit1}
D. Fargion,  {\it  Lett. Nuovo Cim.\ }{\bf 31}, 499 (1981).

\bibitem{gravit2} 
N. Arnaud {\it et al.},  {\it  Phys. Rev.\ } {\bf D65}, 033010 (2002).

\bibitem{steep}
T. Totani,  {\it Phys. Rev. Lett.\ }{\bf  80}, 2039 (1998).

\bibitem{blackhole}
J. F. Beacom, R. N. Boyd and A. Mezzacappa,  {\it Phys. Rev. Lett.\ }{\bf 85}, 3568 (2000);
 {\it Phys. Rev.\ } {\bf D 63}, 073011 (2001).

\bibitem{Nardi:2003pr}
E.~Nardi and J.~I.~Zuluaga,
arXiv:astro-ph/0306384.

\bibitem{neutrino-spectrum}
H. -T. Janka and W. Hillebrandt,
 {\it  Astron. Astrophys.\ } {\bf 224}, 49 (1989).

\bibitem{woosley} 
S. E. Woosley {\it et al.}, 
 {\it  Astrophys. J.\ }  {\bf 433}, 229 (1994).


\end{thebibliography}
\end{document}